\documentclass[12pt,a4paper,oneside]{article}
\usepackage[T1]{fontenc}

\usepackage[utf8]{inputenc}
\usepackage{lmodern}
\usepackage{graphicx}
\usepackage{amsthm}
\usepackage{amssymb}
\usepackage{apacite}    

\usepackage{geometry} %ustalenie formatu papieru
\usepackage{color}
\usepackage{amsmath} %pakiet do składania bardziej skomplikowanych wzorów takich jak ten stronie 3
\font\kapitaliki=plcsc10 % dodefiniowanie nowej czcionki
\def\cR{{\cal R}}
\def\cX{{\cal X}}
\def\Y{{\it YES\ }}

\def\Nd{{\it NO}}

\font\bfgten=cmmib10
\font\bfgseven=cmmib7
\font\bfgfive=cmmib5

\def\bliterka#1{\mathchoice
	{\hbox{{\bfgten#1}}}
	{\hbox{{\bfgten#1}}}
	{\hbox{{\bfgseven#1}}}
	{\hbox{{\bfgfive#1}}}}
\def\blitera#1{\mathchoice
	{\hbox{{\bfgten\char#1}}}
	{\hbox{{\bfgten\char#1}}}
	{\hbox{{\bfgseven\char#1}}}
	{\hbox{{\bfgfive\char#1}}}}

\def\btau{\blitera{28}}
\def\bvarepsilon{\blitera{34}}

\def\bx{\bliterka{x}}
\def\bA{\bliterka{A}}

\title{Robustness against lying of the confidence interval \\ for a fraction of population  with a sensitive characteristic\\}
\date{}

\author{
Wojciech Zieli\'{n}ski\footnote{{\kapitaliki Corresponding author}: Wojciech Zieli\'{n}ski, Department of Econometrics and Statistics, Warsaw University of Life Sciences, Nowoursynowska 159, PL-02-767 Warsaw}\\
Warsaw University of Life Sciences (Poland)\\
e-mail: wojciech$\_$zielinski@sggw.edu.pl\\	}	

\begin{document}
\maketitle
\begin{abstract}
	{\noindent In the investigations devoted to sensitive questions the problem is with the truthfulness of survey participants. We offer a fresh perspective on respondent deception. We investigate the influence of lying on the properties of the confidence intervals for the fraction of the sensitive part of a population.  It is shown that the coverage probability is robust against lying. \\
	}
\end{abstract}
     {\kapitaliki Keywords}:  sensitive questions, robustness, confidence interval\\
     {\kapitaliki 2010 Mathematics Subject Classification}:  62F35 62P25\\
	
\section{Introduction}

We consider the estimation of the percentage of the population who have committed socially stigmatizing misdeeds such as corruption, tax fraud, undeclared work (black market), drug use, violence against
children and other sensitive or illicit activities.

Suppose that the contribution of a respondent in a sensitive group is described by a random variable $\xi$ following a Bernoulli distribution 
$$(1, \pi),\ (0, 1-\pi),$$	
where $\pi$ denotes the percentage of respondents with sensitive characteristic. We construct one--sided (right--sided) confidence interval for this parameter. 

Unfortunately, for obvious reasons, this random variable cannot be directly observed. To address this issue, an auxiliary random variable is introduced, which is related to a neutral topic and independent of the sensitive question asked in the survey (e.g., ``Were you born in May?''). This question is also modeled as a Bernoulli r.v. $\eta$ 
$$(1, q),\ (0, 1-q).$$
The neutral question as well as the probability $q$ are chosen by the experimenter. It is assumed that the probability $q$ is known.

In the survey, we observe a random variable $\delta$ following a Bernoulli distribution
$$(1,\varrho_q(\pi)),\ (0,1-\varrho_q(\pi)).\leqno(*)$$
The probability $\varrho_q(\pi)$ depends on the unknown fraction $\pi$ of people in the sensitive group and the properties $q$ of the neutral question. The  mapping $\pi\to\varrho_q(\pi)$ is a one--to--one function. The choice of this function is made by the experimenter. For example, (Yu, Tian \& Tang, 2008), (Tian 2014) considered the so--called crosswise model in which $\varrho_q(\pi)=(2q-1)\pi+(1-q).$ Based on a sample of zero--one answers the exact one--sided confidence interval is constructed (Clopper \& Pearson 1934)
$$P_\pi\left\{\varrho_q(\pi)\leq\beta^{-1}\left(x+1,n-x;\gamma\right)\right\}\geq\gamma,\leqno{(**)}$$
were $\beta^{-1}\left(a,b;q\right)$ is the $q$-quantile of a Beta distribution with parameters $(a,b)$.

Unfortunately, in practice, respondents may still answer dishonestly. To account for potential lying, we propose assuming that the answers of $n$ respondents form a sequence of independent random variables distributed as  
$$(1,\tau_i),\ (0,1-\tau_i).$$
We assume that probabilites $\tau_1,\ldots,\tau_n$ are such that  $\frac{1}{n}\sum_{i=1}^n\tau_i=\varrho_q(\pi)$. It means that respondents may cheat overstating the real value $\varrho_q(\pi)$ (``gasconade'') or understate it (``shame''), but on average they are truthful.
Our  aim is to investigate the influence of lying on the coverage probability by the above c.i. for the true value of 
$\pi$.

In Section 2 an analysed confidence interval will be given. In Section 3 minimal coverage probability under lying will be derived. In Section 4 with the aid of Monte Carlo simulations it will be shown that slight cheating does not have a substantial influence on the coverage probability and hence potential small lies may be omitted in investigation.

A very similar problem appears in the estimation of the probability of success in dichotomous phenomena on the basis of a sequence of experiments, but the probabiltity of success may vary from experiment to experiment. This topic was considered for example by Hoeffding (1955), Mattner \& Tasto (2015), Bancal \& Sekatski (2024), and in biometrical applications by Elandt (1964), Rao (1973). The aspect of robustness examined in this study was previously addressed in R. Zieliński (1978) and W. Zieliński (1991).

In the present paper we confine ourselves to dichotomous questionaries in which a respondent 
gives answers $YES/NO$. Examples of such investigations may be found for example in  (Warner 1965), (Greenberg, Abdul--Ela \& Horvitz 1969), (Yu, Tian \& Tang 2008), (Tian 2014). 
We do not consider investigations with quantitative answers as in~the case of {\sl Item Count Technique} models (see for example (Kowalczyk, Niemiro \& Wieczorkowski 2021)).

\section{One-sided confidence interval}

Our aim is to construct on the basis of a sample of size $n$ right--sided confidence interval for $\pi$. We have a sample $\delta_1,\ldots,\delta_n$ from the distribution $(*)$. We construct a confidence interval for  $\varrho_q(\pi)$ and then using injectivity of the mapping $\pi\to\varrho_q(\pi)$ we obtain a c.i. for $\pi$.

Let $x$ successes (ones) be observed in a sample of size $n$. An exact  one--sided confidence interval at the $\gamma$ confidence level for $\varrho_q(\pi)$ has the form (Clopper \& Pearson 1934)
$$P_\pi\left\{\varrho_q(\pi)\leq\beta^{-1}\left(x+1,n-x;\gamma\right)\right\}\geq\gamma$$
for all $\pi\in(0,1)$. For $x=n$ we take $1$ as a right end. Here $\beta^{-1}\left(a,b;q\right)$ is the  $q$-quantile of a Beta distribution with parameters $(a,b)$. We do not use here a common approximation by Central Limit Theorem (CLT), because obtained intervals do not fulfill the fundamental requirement given by (Neyman 1934, p. 562), so it would be incorrect to call those intervals confidence ones.

\section{Coverage probability}

In the absence of lying, the coverage probability of the confidence interval for $\rho_q(\pi)$ equals
$$Cov_{n,\gamma}(\varrho_q(\pi))= 
\sum_{k=0}^{m}\binom{n}{k}\varrho_q(\pi)^k(1-\varrho_q(\pi))^{n-k},$$
$m=Bin^{-1}_{n,\gamma}(\varrho_q(\pi))$ (here $Bin^{-1}_{n,q}(\theta)$ is the $q$-quantile of Binomial distribution $Bin(n,\theta)$). In what follows we omit $n$ and $\gamma$ treating them as given. 
By construction of the confidence interval $(**)$ it follows that $Cov_{n,\gamma}(\varrho_q(\pi))\geq\gamma$ for all $\pi\in(0,1)$. 

Assume now that real probabilities of the answer \Y are equal $\btau=(\tau_1,\ldots,\tau_n)$, i.e. are biased with lying. We accept that we have $n$ zero--ones answers, i.e. we assume that we observe $n$ independent r.v.'s 
distributed as Bernoulli $Ber(\tau_i), i=1,\ldots,n$, where $$\tau_i=\varrho_q(\pi)+\varepsilon_i,\ \sum_{i=1}^{n}\varepsilon_i=0.$$ This last assumption stems from the observation that some respondents tend to overreport (out of pride), while others underreport (due to embarrassment or fear of stigma). Assume that on average effects of lying cancel out.

Let $$\cR(\pi)=\left\{\btau=(\tau_1,\ldots,\tau_n): \sum \tau_i=n\varrho_q(\pi), \tau_1,\ldots,\tau_n\in[0,1]\right\}$$
be the set of all distributions disturbed by lies. Consider also the set of all small lies, i.e. 
$$\cR^{\delta}(\pi)=\left\{\btau=(\tau_1,\ldots,\tau_n): \sum \tau_i=n\varrho_q(\pi), \tau_1,\ldots,\tau_n\in[0,1],|\tau_i-\varrho_q(\pi)|<\delta\right\},$$ where $\delta$ is the given level of small lies. 

Let $$\cX^n(k)=\left\{\bx=(x_1,\ldots,x_n): \sum x_i=k, x_1,\ldots,x_n\in\{0,1\}\right\}$$ 
be the set of all answers from $n$ questionnaires containing exactly $k$ \Y responses.

Let  $\bvarepsilon=\btau-\varrho_q(\pi)$ be the vector of probabilities in the presence of lies. Now the probability of coverage of the value $\varrho_q(\pi)$ equals
$$Cov_{n,\gamma}(\varrho_q(\pi),\btau)= 
\sum_{k=0}^{m}\sum_{\bx\in\cX(k)}
\tau_1^{x_1}(1-\tau_1)^{1-x_1}\cdots \tau_n^{x_n}(1-\tau_n)^{1-x_n}.$$
Let $w(\tau_j)=\tau_j^{x_j}(1-\tau_j)^{1-x_j}$. The coverage probability by c.i. $(**)$ is a function
$$(\varrho_q(\pi),\btau)\to Cov_{n,\gamma}(\varrho_q(\pi),\btau)= 
\sum_{k=0}^{m}\sum_{\bx\in\cX(k)}\prod_{j=1}^nw(\tau_j).$$
We are interested in the function
$$\varrho_q(\pi)\to \inf_{\btau\in\cR(\pi)}Cov_{n,\gamma}(\varrho_q(\pi),\btau).$$
The function $Cov_{n,\alpha}(\varrho_q(\pi),\btau)$ may be written in one of following equivalent forms
$$\begin{aligned}
	Cov_{n,\gamma}(\varrho_q(\pi),\btau)&= 
	\sum_{k=0}^{m}\sum_{\bx\in\cX^n(k)}\prod_{j=1}^nw(\tau_j)\cr 
	&=\sum_{k=0}^{m}\left\{\tau_1\sum_{\bx_{(1)}\in\cX^{n-1}(k-1)}\prod_{j=1\atop j\neq 1}^nw(\tau_j)
	+(1-\tau_1)\sum_{\bx_{(1)}\in\cX^{n-1}(k)}\prod_{j=1\atop j\neq 1}^nw(\tau_j)\right\}\cr
	&=\cdots=\cr
	&=\sum_{k=0}^{m}\left\{\tau_n\sum_{\bx_{(n)}\in\cX^{n-1}(k-1)}\prod_{j=1\atop j\neq n}^nw(\tau_j)
	+(1-\tau_n)\sum_{\bx_{(n)}\in\cX^{n-1}(k)}\prod_{j=1\atop j\neq n}^nw(\tau_j)\right\}\cr
\end{aligned}
$$
where $\bx_{(i)}=\bx-x_i$. The symbol $\bx_{(i)}$ means an $(n-1)$-vector $(x_1,\ldots,x_{i-1},x_{i+1},\ldots,x_n)$. We take $\cX^{n-1}(-1)=\emptyset$.

We will find extreme points of $Cov_{n,\gamma}(\varrho_q(\pi),\btau)$ in the set $\cR(\pi)$, i.e. under condition $\sum \tau_i=n\varrho_q(\pi)$. The method of Lagrange multipliers gives a system of $n+1$ equations (with a multiplier~$\lambda$)
$$\begin{cases}
	\displaystyle\sum_{k=0}^{m}\left(\sum_{\bx_{(i)}\in\cX^{n-1}(k-1)}\prod_{j=1\atop j\neq i}^nw(\tau_j)
	-\sum_{\bx_{(i)}\in\cX^{n-1}(k)}\prod_{j=1\atop j\neq i}^nw(\tau_j)\right)-\lambda=0,&i=1,\ldots,n,\cr
	\sum \tau_i-n\varrho_q(\pi)=0.&\cr
\end{cases}$$
After some simplifications we obtain
$$\begin{cases}
	\displaystyle\sum_{\bx_{(i)}\in\cX^{n-1}(m)}\prod_{j=1\atop j\neq i}^nw(\tau_j)+\lambda=0,&i=1,\ldots,n,\cr
	\sum \tau_i-n\varrho_q(\pi)=0.&\cr
\end{cases}$$
Subtracting the second, third, ..., and $n$-th equations from the first, we obtain the following system of $n$ equations with unknowns $\tau_1,\ldots,\tau_n$:
$$\begin{cases}
	(\tau_1-\tau_2)\left\{\displaystyle\sum_{\bx_{(1,2)}=m-2}\prod_{j\neq1,2}w(\tau_j)-\sum_{\bx_{(1,2)}=m-1}\prod_{j\neq{1,2}}w(\tau_j)\right\}=0\cr
	\vdots\cr
	(\tau_1-\tau_n)\left\{\displaystyle\sum_{\bx_{(1,n)}=m-2}\prod_{j\neq1,n}w(\tau_j)-\sum_{\bx_{(1,n)}=m-1}\prod_{j\neq1,n}w(\tau_j)\right\}=0,\cr
	\sum \tau_i-n\varrho_q(\pi)=0,\cr
\end{cases}$$
where $\bx_{(i,j)}=\bx-x_i-x_j$.

The above system has a solution $\tau_1=\cdots=\tau_n=\varrho_q(\pi)$. This is a situation when respondents are truthful i.e. they do not lie.

Let $\bA(\btau)=[a_{ij}(\btau)]_{i,j=i,\ldots,n}$ be the matrix of second partial derivatives of $Cov_{n,\gamma}(\varrho_q(\pi),\btau)$: 
$$\left.\frac{\partial^2Cov_{n,\gamma}(\varrho_q(\pi),\btau)}{\partial\tau_i\partial\tau_j}\right|_{i,j=1,\ldots,n}$$
This matrix is as follows:
$$a_{ij}(\btau)=\begin{cases}
	0,&\hbox{for }i=j,\cr
	\displaystyle\sum_{x_{(i,j)}=k-2}\prod_{m\neq i,j}w(\tau_m)-\sum_{x_{(i,j)}=k-1}\prod_{m\neq i,j}w(\tau_m),&\hbox{for }i\neq j.\cr
\end{cases}$$
Let $\bar\btau$ denote the vector of $\tau_i= \varrho_q(\pi),\ i=1,\ldots,n$.
For $\bar\btau$ we obtain
$$a_{ij}(\bar{\btau})=
\begin{cases}
	0,&\hbox{for }i=j\cr
	c,&\hbox{for }i\neq j,\cr
\end{cases}$$
where
$$c={n-2\choose k-2}\varrho_q(\pi)^{k-2}(1-\varrho_q(\pi))^{n-k-1}\left(1-{n-1\over k-1}\varrho_q(\pi)\right).$$
The major $m$ minor of the matrix $\bA(\bar{\btau})$ equals $(m-1)(-1)^{m-1}c^m$ (Rao 1973, p. 67). In particular, the determinant of the matrix $\bA(\bar{\btau})$ equals $(n-1)(-1)^{n-1}c^n$.

Consecutive major minors of the matrix $\bA(\bar{\btau})$ are of the oposite signs, hence the point  $\bar{\btau}$ is a local maximum of the function $Cov_{n,\gamma}(\varrho_q(\pi),\btau)$:
$$\sup_{\cR(\pi)}Cov_{n,\gamma}(\varrho_q(\pi),\btau)=Cov_{n,\gamma}(\varrho_q(\pi),\bar{\btau})$$
It follows that $\inf_{\btau\in\cR(\pi)}Cov_{n,\gamma}(\varrho_q(\pi),\btau)$ is obtained on the edge of the cube $\cR(\pi)$. To estimate the value of that infimum Monte--Carlo method will be aplied (see the next Section).
%\bigskip

If we confine ourselves to ``small'' lies, i.e. if we suppose that the true probabilites are in the set  $\cR^{\delta}(\pi)$ the following proposition may be proved
\bigskip

\noindent{\bf Proposition.} For $(\tau_1,\ldots,\tau_n)\in \cR^{\delta}(\pi)$ it occurs
$$|Cov_{n,\gamma}(\varrho_q(\pi),\btau)-Cov_{n,\gamma}(\varrho_q(\pi))|\leq C\delta^2,$$
where $C$ is a constant.
\smallskip

\noindent{\it Proof}. 
A second-order Taylor expansion of the function $Cov_{n,\gamma}(\varrho_q(\pi),\btau)$ around the point $\varrho_q(\pi)$ yields:
$$\begin{aligned}
	Cov_{n,\gamma}(\varrho_q(\pi),\btau)&=Cov_{n,\gamma}(\varrho_q(\pi))+{n-1\choose k-1}\varrho_q(\pi)^{k-1}(1-\varrho_q(\pi))^{n-k}\left(\sum \tau_i-n\varrho_q(\pi)\right)+\cr
	&\frac{1}{2}\sum_{i,j}\frac{\partial^2}{\partial \tau_i\partial \tau_j}Cov_{n,\gamma}(\tau_1^0,\ldots,\tau_n^0)(\tau_i^0-\varrho_q(\pi))(\tau_j^0-\varrho_q(\pi)),
\end{aligned}$$
where $(\tau_1^0,\ldots,\tau_n^0)\in \cR^{\delta}(\pi)$. Since $(\tau_1,\ldots,\tau_n)\in \cR^{\delta}(\pi)$, hence $\sum \tau_i-n\varrho_q(\pi)=0$. The second derivative of the function $Cov_{n,\gamma}(\varrho_q(\pi),\btau)$ equals
$$\frac{\partial^2}{\partial \tau_i\partial \tau_j}Cov_{n,\gamma}(\varrho_q(\pi),\btau)=
\begin{cases}
	0,&\hbox{for $i=j$},\cr
	\displaystyle\sum_{x_{(i,j)}=k-2}\prod_{m\neq i,j}w(\tau_m)-\sum_{x_{(i,j)}=k-1}\prod_{m\neq i,j}w(\tau_m),&\hbox{for $i\neq j$}\cr
\end{cases}$$
is uniformly bounded on the set $\cR^{\delta}(\pi)$ by some constant $M$:
$$\left|{\partial^2\over\partial \tau_i\partial \tau_j}Cov_{n,\gamma}(\varrho_q(\pi),\btau)\right|\leq M,\quad\left(\forall i,j\right).$$
Hence
$$\begin{aligned}
	&\left|Cov_{n,\gamma}(\varrho_q(\pi),\btau)-Cov_{n,\gamma}(\varrho_q(\pi))\right|=\cr
	&{1\over2}\left|\sum_{i,j}{\partial^2\over\partial \tau_i\partial \tau_j}Cov_{n,\gamma}(\varrho_q(\pi),\btau^0)(\tau_i^0-\varrho_q(\pi))(\tau_j^0-\varrho_q(\pi))\right|\leq{n(n-1)\over2}M\delta^2.
\end{aligned}$$
Taking $C=n(n-1)M/2$ we obtain the thesis. $\blacksquare$ %$\square$

\bigskip

From the proven fact we conclude that little lies are not significant for the probability of the coverage. In that sense confidence interval is robust against described little lies. This means that you do not have to worry too much about the fact that respondents may give slightly false answers. It must be stressed that our conclusions do not depend on the model applied in research.

%\bigskip

\section{Monte--Carlo investigations}

Analytical calculation of deviations from the nominal confidence level is a formidable task and, in fact, appears to be intractable. In what follows to estimate those deviations computer simulations will be applied. For further research we choose the crosswise model of questionnaire (Yu, Tian \& Tang 2008).

In that model, along with a sensitive question, a neutral one is asked. This question is a dichotomous one with the probability $q\in(0,0.5)$. The response to the questionnaire equals $1$ if a respondent in the sensitive group answers \Y to the neutral question, or if a respondent in the non-sensitive group answers \Nd. Otherwise the response to the whole questionnaire equals $0$. So we observe a Bernoulli random variable 
$$Z=\begin{cases}
	1,&\hbox{with probability $\varrho_q(\pi)=q\pi+(1-q)(1-\pi)$}, \\
	0,&\hbox{with probability $1-\varrho_q(\pi)$.} \\
\end{cases}$$	
We draw a sample $u_1,\ldots,u_n$ from the uniform distribution $U(0,1)$. Those numbers will represent answers to the questionnaire: $z_i=1$, if $u_i<\varrho_q(\pi)$, and $z_i=0$ otherwise. In that manner we obtain a sample of reliable answers. Now we draw a sample of lies, i.e. the sample $t_1,\ldots,t_n$ independent random variables uniformly distributed $U(-\delta,\delta)$. Number $\delta$ may be interpreted as the maximal level of cheating. Answers in the questionnarie are biased with lies: $z_i=1$ if $u_i<\varrho_q(\pi)+t_i$ and $z_i=0$ otherwise. Random numbers $t's$ are slightly modified in such a way they sum up to zero: $t_i:=t_i-\bar t$.

In simulations we take $10000$ respondents. As a confidence level we adopt $0.95$. Due to the fact that confidence interval is built for $\varrho_q(\pi)$ and we are interested in the estimation of $\pi$ i.e. a sensitive percentage we have to recalculate parameters on $\pi$. In Table~1 the dependence of $\varrho_q(\pi)$ on parameters of the model is shown.

\noindent 	
\begin{center}
	\textbf{Table 1:} Parameters applied in simulations\\ 
	\begin{small}
		\begin{tabular}{ccccc}\hline
			&$q=0.1$&$q=0.2$&$q=0.3$&$q=0.4$\\ \hline
			$\pi=0.01$&0.892&0.794&0.696&0.598\\			
			$\pi=0.05$&0.86&0.77&0.68&0.59\\		
			$\pi=0.10$&0.82&0.74&0.66&0.58\\			
			$\pi=0.15$&0.78&0.71&0.64&0.57\\ \hline	 		
		\end{tabular}
	\end{small}
\end{center}

In simulations we take $0.5\leq\varrho_q(\pi)\leq0.9$. That range covers, from the practical point of view, all interesting situations.

In Table~2 the simulated coverage probabilities are presented. In the column $\delta=0$ there are given coverage probabilities for truthful respondents. Note that a larger parameter $\delta$ indicates a stronger tendency toward untruthful responding.  

\noindent 	
\begin{center}
	\textbf{Table 2:} Coverage probabilities\\ \smallskip
	\begin{small}
		\begin{tabular}{ccccc}\hline
			&$\delta=0$&$\delta=0.01$&$\delta=0.02$&$\delta=0.03$\\ \hline
			$\varrho_q(\pi)=0.892$&0.9543&0.9533&0.9546&0.9526\\
			$\varrho_q(\pi)=0.794$&0.9545&0.9546&0.9530&0.9550\\
			$\varrho_q(\pi)=0.696$&0.9569&0.9558&0.9557&0.9581\\
			$\varrho_q(\pi)=0.598$&0.9591&0.9592&0.9598&0.9585\\
			$\varrho_q(\pi)=0.500$&0.9575&0.9551&0.9554&0.9542\\ \hline
		\end{tabular}
	\end{small}
\end{center}

Since the simulated coverage probability  does not differ significantly from the nominal confidence level we may conclude that confidence level is robust against lying.

\section{Conclusions}

This study addresses the problem of estimating the proportion of a population possessing a sensitive characteristic. However, in real-world applications, data quality is severely compromised by response bias. To collect appropriate data it is rather impossible to ask directly about sensitive characteristic and one has camouflage this question. The first method of hiding sensitive question was proposed by (Warner 1965). Since that seminal paper, numerous other models have been proposed in the literature. In our paper we confine ourselves to the crosswise model proposed by (Tian, Yu, Tang \& Geng 2007). 

Nevertheless, despite efforts to ensure anonymity, some respondents will lie. A mathematical model of the phenomena of lying in the answers to a questionnaire was proposed. In the paper we analyse the effect of lying on the statistical properties of the confidence interval for the sensitive fraction of a population. On the basis of computer simulations it is shown confidence level is robust to lies.

Those results should be treated as preliminary. In future research, alternative models will be analyzed, including the Item Count Technique model  (see for example (Kowalczyk, Niemiro \& Wieczorkowski 2021)).

It was assumed that the effects of ``gasconade'' and ``shame'' are of the same importance. But in some real problems one of those effects may appear more frequently. The ``assymetry'' of lying will be analysed as well.

\begin{center}
	\includegraphics[scale=0.8]{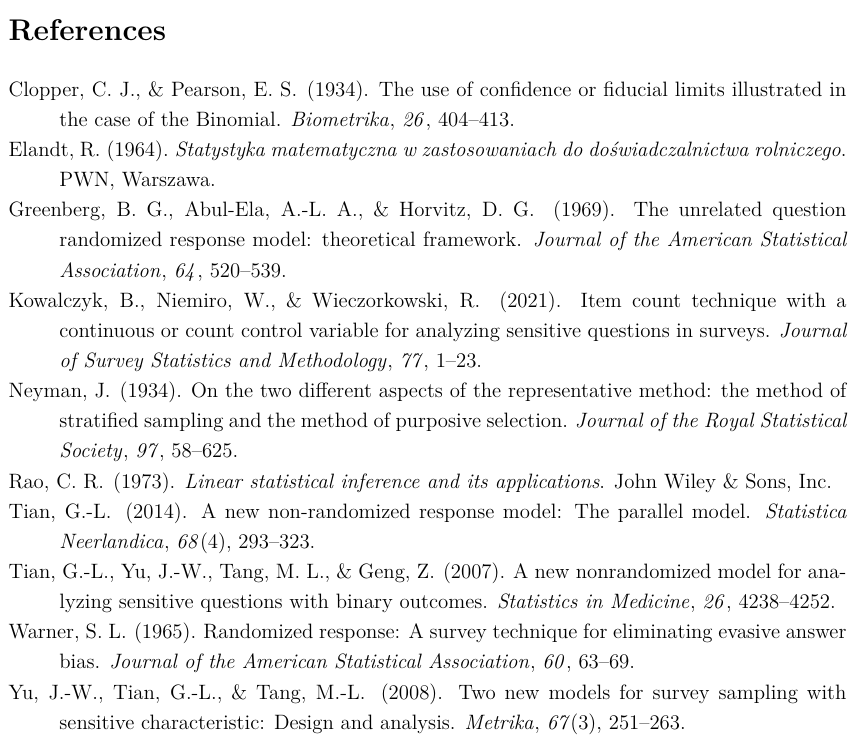}
	%		\caption{Length versus $\pi$ for different $\lambda$ ($m=150$). }
	%		\label{graph:m150}
\end{center}

\begin{center}
	\includegraphics[scale=0.8]{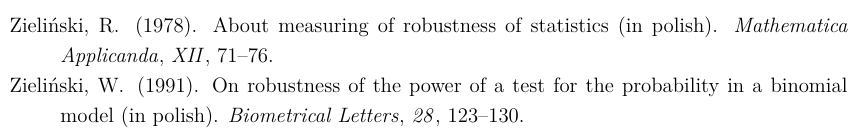}
	%		\caption{Length versus $\pi$ for different $\lambda$ ($m=150$). }
	%		\label{graph:m150}
\end{center}

\end{document}